\documentclass[useAMS,usegraphicx,usenatbib]{mn2e}

\usepackage{graphicx}
\usepackage{times}
\usepackage{amssymb}
\usepackage{amsmath}
\usepackage{euscript}
\usepackage{longtable,lscape}

\def\210keV{{\rm\thinspace 2--10 keV}}
\usepackage{wasysym}
\title[Spinning black holes and the X-ray Background]{A selection effect boosting the contribution from rapidly spinning black holes to the Cosmic X-ray Background}
\author[R.V. Vasudevan et al.]{R.V. Vasudevan$^1$\thanks{e-mail:ranjan@ast.cam.ac.uk}, A.C. Fabian$^1$, C. S. Reynolds$^{1,2}$, J. Aird$^{1}$, T. Dauser$^{3}$ and L.C.Gallo$^{4}$\\\footnotesize$^1$ Institute of Astronomy, Madingley Road, Cambridge CB3 0HA \\\footnotesize$^2$ Department of Astronomy, University of Maryland, College Park, MD 20742, USA\\\footnotesize$^3$ Dr Karl Remeis-Observatory and Erlangen Centre for Astroparticle Physics, Sternwartstr. 7, D-96049 Bamberg, Germany\\\footnotesize$^4$ Department of Astronomy and Physics, Saint Mary's University, 923 Robie Street, Halifax, Nova Scotia B3H 3C3, Canada}

\begin{document}

\maketitle
\begin{abstract}
The Cosmic X-ray Background (CXB) is the total emission from past accretion activity onto supermassive black holes in active galactic nuclei (AGN) and peaks in the hard X-ray band (30~keV).  In this paper, we identify a significant selection effect operating on the CXB and flux-limited AGN surveys, and outline how they must depend heavily on the spin distribution of black holes.  We show that, due to the higher radiative efficiency of rapidly-spinning black holes, they will be over-represented in the X-ray background, and therefore could be a dominant contributor to the CXB.  Using a simple bimodal spin distribution, we demonstrate that only 15~per~cent maximally-spinning AGN can produce 50~per~cent of the CXB.  We also illustrate that invoking a small population of maximally-spinning black holes in CXB synthesis models can reproduce the CXB peak without requiring large numbers of Compton-thick AGN.  The spin bias is even more pronounced for flux-limited surveys: 7~per~cent of sources with maximally-spinning black holes can produce half of the source counts.  The detectability for maximum spin black holes can be further boosted in hard ($>$10~keV) X-rays by up to $\sim$ 60~per~cent due to pronounced ionised reflection, reducing the percentage of maximally spinning black holes required to produce half of the CXB or survey number counts further.   A host of observations are consistent with an over-representation of high-spin black holes. Future NuSTAR and ASTRO-H hard X-ray surveys will provide the best constraints on the role of spin within the AGN population.

\end{abstract}

\begin{keywords}
black hole physics -- galaxies: active  -- quasars: general -- galaxies: Seyfert
\end{keywords}

\section{Introduction}
\label{Intro}

The cosmic X-ray background (CXB) is the combined emission from accretion onto SMBHs in active galactic nuclei (AGN).  Whilst the basic X-ray AGN spectrum is a power-law from Comptonization in the corona, the spectrum of each AGN is modified further by differing degrees of absorption by neutral and/or ionised gas and Compton reflection from both distant, neutral material and from the accretion disc itself.   Absorption reduces the flux at lower energies, causing a rising spectrum towards 10~keV, and reflection boosts the flux above 10~keV.  These effects combine in different ways to produce the observed CXB spectrum.  However there is a high degree of uncertainty as to exactly what combination of these effects is required to produce the CXB spectrum.  \cite{1989A&A...224L..21S} realised early on that absorption in the AGN population can be invoked to explain the CXB; \cite{1990MNRAS.242P..14F} first proposed using reflection-dominated sources to explain the shape of the CXB.  Earlier CXB synthesis models required a fraction of Compton-thick AGN up to $\sim$30 per cent \citep{2007A&A...463...79G}, whereas later studies (e.g., \citealt{2009ApJ...696..110T}, \citealt{2012A&A...546A..98A}) showed that the reflection contribution to the 30~keV peak of the CXB can be increased at the expense of the Compton-thick fraction to successfully explain the CXB spectrum.  The relative importance of reflection and absorption have significant implications for what the true energy budget is contributing to the CXB: is most of it enshrouded and appears re-processed by dust in the infrared, or is what we are seeing the true emission from the central engine?

A key factor in determining the energy budget of the CXB is the
radiative efficiency of accretion onto black holes, an issue which is
relevant for both reflection-dominated and absorption-dominated
AGN. Accretion onto supermassive black holes is the second most
efficient energy-generation mechanism known in the Universe, after
matter-antimatter annihilation.  The efficiency of this process is
increased around a rapidly spinning black hole, as the smaller event
horizon and innermost stable circular orbit (ISCO) allow gravitational
energy to be extracted from deeper into the steep potential well: for
a black hole at the maximal spin allowed (i.e., $a_{\rm max}$=0.998 when taking torque at the inner edge of the disc into account, \citealt{1974ApJ...191..507T}), the event horizon and ISCO are 1.06 and
1.24 $R_{\rm g}$, compared to 2 and 6 $R_{\rm g}$ for a non-spinning
black hole (where $R_{\rm g}=GM/c^{2}$ is the gravitational radius for a
black hole of mass M).  \cite{1974ApJ...191..507T} shows that half of
the radiation from a non-spinning black hole emerges from within 30
gravitational radii ($R_{\rm g}$), but for a rapidly spinning black
hole, half of the radiation emerges from within 5$R_{\rm g}$ (see also
Fig.~1 of \citealt{2000ApJ...528..161A}).  

For high spin, we can therefore probe deeper into the accretion flow, and it is important to
understand whether the CXB places any constraints on where the
emission is coming from in a global, average sense.  X-rays from the
corona illuminate the disc and produce a scattered (reflected)
spectrum from the inner accretion disc.  The precise shape of the X-ray reflection spectrum is a powerful probe of spin, and the position of the X-ray
source, as revealed by the strength of the reflected spectrum, can
provide a powerful probe of the geometry of the system, since rapidly spinning black holes
have a smaller ISCO.  The X-ray source can therefore be located even
closer to the black hole in such sources, allowing more pronounced
General Relativistic effects to be seen.  Assuming the standard
geometry of a compact corona at height $h$ above the accretion disc
poles (a `lamp-post' geometry), the strength of the features in the
reflection spectrum and the degree of relativistic smearing, provide a
measure of $h$, and therefore constrain the black hole spin (\citealt{1996MNRAS.282L..53M}, \citealt{2013MNRAS.428.2901W}, \citealt{2013MNRAS.430.1694D}). Such reflection model-fitting generally yields steep emissivity profiles, implying a low source height.  These findings, together with X-ray reverberation studies (\citealt{2013MNRAS.436.3782D};
\citealt{2013MNRAS.428.2795K}, \citealt{2013MNRAS.434.1129K};
\citealt{2014MNRAS.439.3931E}; see \citealt{2014A&ARv..22...72U} for a
review) and estimates of source compactness \citep{2015arXiv150507603F}, $h$ is constrained to lie in the range of $2-10 r_{\rm g}$.
Indeed, it is plausible that $h\sim2-7r_{\rm ISCO}$, where $ r_{\rm
  ISCO}$ is the radius of the ISCO. A theoretical study on reverberation with a vertically-extended corona by \cite{2016arXiv160200022W} suggests that part of the corona could be located as low as 1.5 $R_{\rm g}$, to explain features seen in recent timing studies.

The spin distribution of supermassive black holes has important wider relevance in understanding galaxy evolution. The final spin distribution of SMBHs is set by the accretion and merger history of galaxies (\citealt{2013ApJ...775...94V}; \citealt{2014ApJ...794..104S}, \citealt{2013ApJ...762...68D}), so can act as a probe of inner galaxy evolution. High spin indicates more continuous accretion with a similar angular momentum as compared to low spin which points to chaotic accretion \citep{2006MNRAS.373L..90K}.

\begin{figure}
\centerline{
\includegraphics[width=8.0cm]{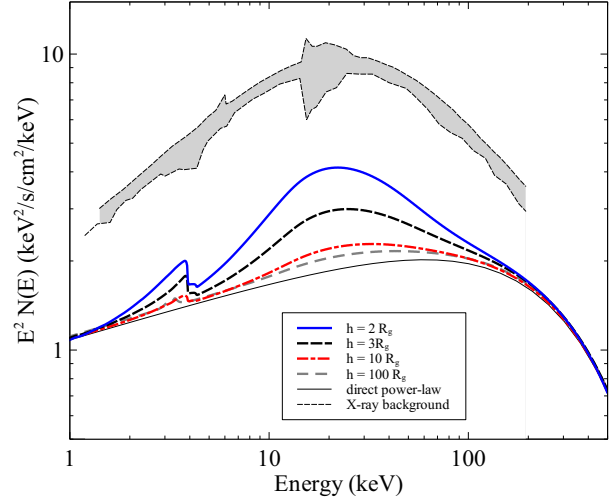}
}
\caption{Illustration of how bringing the X-ray source closer to the black hole can produce a spectrum of similar shape to the Cosmic X-ray Backgroud (CXB) peak due to light bending (e.g., \protect\citealt{2007MNRAS.382.1005G}), assuming an inclination of $i=60^{\circ}$ and redshift $z=1$. An ionisation parameter of $\xi = L/nR^{2}$ of 50 has been adopted ($L$ is the source luminosity, $n$ the density of the material and $R$ the distance of the material from the source of ionising radiation).  The CXB spectrum (dashed line with grey shaded area showing uncertainty) is from \protect\cite{2009A&A...493..501M}.
\label{relxill_cxb_comparison}}
\end{figure}

The most direct method of determining spin in AGN is via relativistic
blurring of the iron K$\alpha$ line at 6.4~keV, and while spin has
been thus determined for a few tens of objects, such spin estimates
are not yet available for large, flux-limited samples of AGN due to
the quality of data required. If the reflection is occurring in the
inner parts of the accretion flow, the shape of the hard X-ray excess
above 10~keV will be more peaked for X-ray sources closer to the black
hole, producing a shape similar to the CXB peak
(Fig.~\ref{relxill_cxb_comparison}).

The luminous output from accretion is given by
$L = \eta \dot{M} c^{2}$ for radiative accretion efficiency $\eta$, mass
accretion rate $\dot{M}$ and speed of light $c$.  The radiative accretion efficiency
can range from $\eta = 0.057$ for a non-spinning black hole to $0.32$
for maximal spin ($a=0.998$), so for a given distribution of mass accretion rates,
rapidly spinning black holes should be more frequently sampled by
surveys and also contribute more to the cosmic X-ray background due to
their greater luminosities.  In particular, the steep increase of
$\eta$ with spin at the very highest spins will act to over-represent
the highest spins even more. It is clear that for any realistic intrinsic spin-distribution (i.e., one which consists of more than a single spin), this effect will be at work: if there is even a small spread in intrinsic spins, the higher spins will be over-represented due to their higher efficiency. 

In this work, we aim to adress the question of how this bias towards high spin affects our understanding of the composition of the CXB, flux-limited surveys and luminosity functions.  We make the assumption that there is indeed a spread of spin values in the real AGN population (justified by the observed spin distribution measured so far for AGN). The presence of this selection effect for spin has been briefly explored before by
\cite{2011ApJ...736..103B}; there the emphasis was on flux-limited surveys which are particularly sensitive to spin.  We quantify this selection effect further in
this paper, exploring the effects of spin in the Cosmic X-ray Background (CXB).


This paper is organised as follows: in \S\ref{effcorr} we discuss the role of radiative efficiency in boosting the contribution from high-spin in the CXB.  In \S\ref{fluxlimsurv} we consider how flux-limited surveys would be affected by the effects of efficiency boosting for high spin, including a consideration of X-ray band-specific effects.  In \S\ref{highspinsamples} we discuss other independent evidence for an over-representation of high-spin in AGN studies to-date. In \S\ref{othereff} we identify further effects for consideration in future works, in \S\ref{summary} we summarise our work and identify some consequences of the high spin boost, and in Appendix~\ref{app1} we collate a host of observational results consistent with the over-representation of high spin.

\section{X-ray background modelling: accretion efficiency corrections}
\label{effcorr}

The over-representation of rapid spin may have a direct bearing on the spectral make-up of the CXB.  As recent CXB models find that reflection can play a more important role in accounting for the CXB, and if by `natural selection' a large proportion of black holes contributing to the CXB are rapidly spinning, the reflection will be mostly ionised and from the inner parts of the accretion flow.  The observed reflection fractions from surveys ($R \approx 2$, e.g., \citealt{2015arXiv151104184A}, \citealt{2013ApJ...770L..37V}) have been explained as due to distant-reflection accompanying moderate levels of absorption, but these $R$ values may be alternatively explained by invoking strong light-bending close to a spinning black hole.
  

The role of reflection in CXB synthesis has been extensively discussed previously: \cite{1990MNRAS.242P..14F} showed that the CXB could be composed of reflection-dominated sources generating a hard spectrum above 10~keV. \cite{1995ApJ...438L..63Z} suggested that significant reflection was not required to fit the CXB based on an average 2-500~keV AGN spectrum from \emph{Ginga} and \emph{OSSE} data, but subsequent X-ray background synthesis models have favoured including reflection, including \cite{2007MNRAS.382.1005G}, \cite{2009ApJ...696..110T} and \cite{2012A&A...546A..98A}. \cite{2007MNRAS.382.1005G} particularly discuss how constraints can be placed on the percentage of light-bent sources (implying ionised, inner-accretion-flow reflection) that can be invoked to account for the X-ray background.  They predicted, at a time when \emph{Swift} and \emph{INTEGRAL} were beginning to provide observations of large samples of AGN at the peak energy of the CXB, that it is difficult to reproduce the X-ray background spectrum without invoking reflection.  If one allows for part of the hard excess to be produced by distant reflection, there is still freedom for up to 50 per cent of sources to have significant light-bending producing the hard excess, implying rapidly spinning black holes.  \cite{2007MNRAS.382.1005G} find that if the X-ray source height is small (as low as 2 $R_{\rm g}$), the proportion of light-bent sources required to fit the XRB is lower (5 per cent), since light bending enhances the peak at 30~keV.  Since a variety of source heights are likely, the true fraction of light-bent sources probably lies between these extremes.  If the X-ray source is located away from the accretion disc rotation axis, Doppler boosting can further enhance the X-ray peak \citep{2006A&A...453..773S}.

Additionally, \cite{2004MNRAS.351..169M} find that local black holes can be completely accounted for as relics of past AGN activity, if the accretion process is efficient ($\eta = 0.04-0.16$) and happens at high Eddington ratios (0.1--1.7).  This strengthens the early conclusions of \cite{1982MNRAS.200..115S}, who argue for high radiative efficiency in the CXB based on AGN number counts, implying high spin on average.   They state that the black hole mass density in the local Universe $\rho_{\rm BH}$ can be directly related to the luminous energy density $\epsilon$ seen in AGN surveys or the CXB as follows:

\begin{equation} \label{eq:soltan}
(1 + \langle z \rangle) \epsilon = \frac{\eta}{1-\eta} \rho_{\rm BH} c^{2},
\end{equation}
where $\langle z \rangle$ is the average redshift at which the accretion took place. The factor $\eta/(1-\eta)$ appears because for a quantity of mass-energy $mc^{2}$ falling onto the black hole, $\eta mc^{2}$ is radiated away and the remainder $(1-\eta)mc^{2}$ is accreted onto the black holes to add to their mass.  The energy $\eta mc^{2}$ is what appears as luminous output to the observer at infinity.  

We can use this simple formalism to show how populations of black holes with different spins can contribute at different levels to the overall CXB flux. We can write a version of equation~(\ref{eq:soltan}) that links the emitted radiation not to the final mass density of black holes $\rho_{BH}$, but that links to the total accreted mass density $\rho_{\rm acc}$:

\begin{equation} \label{eq:soltan_rad}
(1 + \langle z \rangle) \epsilon = \eta \rho_{\rm acc} c^{2},
\end{equation}
which can be understood as a global re-statement of $L=\eta \dot{M} c^{2}$ for the entire AGN population.  This expression links the radiative output from black holes to the accretion rate distribution in the population via $\rho_{\rm acc}$.  We can now quantify the degree to which spin can affect the CXB using the dependence of accretion efficiency $\eta$ on spin $a$.  Consider an AGN population with spin $a_{i}$ in a given redshift shell.  We assume that the feeding mechanism is independent of spin, i.e., that $\dot{M}$ for any individual object (and therefore $\rho_{\rm acc}$ for the population) are independent of spin.  The contribution of this population to the X-ray background flux in this redshift shell will be boosted by the efficiency factor $\eta$. The contribution of a population of AGN with spin $a_{i}$ is then related to the fraction of the population at that spin $f(a_{i})$, modulated by the efficiency:

\begin{equation} \label{eq:fracduetospini}
F(a_{i}) \propto \eta_{i} f(a_{i})
\end{equation}

As an illustrative example, we consider a simple two-population model with $a_{1}=0.0$ and $a_{2} = 0.5, 0.8, 0.95, 0.998$.  The fraction of the total population due to high spin is $f(a_{2})$, and the fraction due to low spin is $f(a_{1}) = 1 - f(a_{2})$.  The fraction of flux produced by the high-spin population can straightforwardly be determined by normalising the distribution in equation (\ref{eq:fracduetospini}):

\begin{equation} \label{eq:fracduetohighspin}
F(a_{2}) = \frac{f(a_{2}) \eta_{2}}{[1-f(a_{2})] \eta_{1} + f(a_{2}) \eta_{2}}
\end{equation}

\begin{figure}
\centerline{
\includegraphics[width=8.0cm]{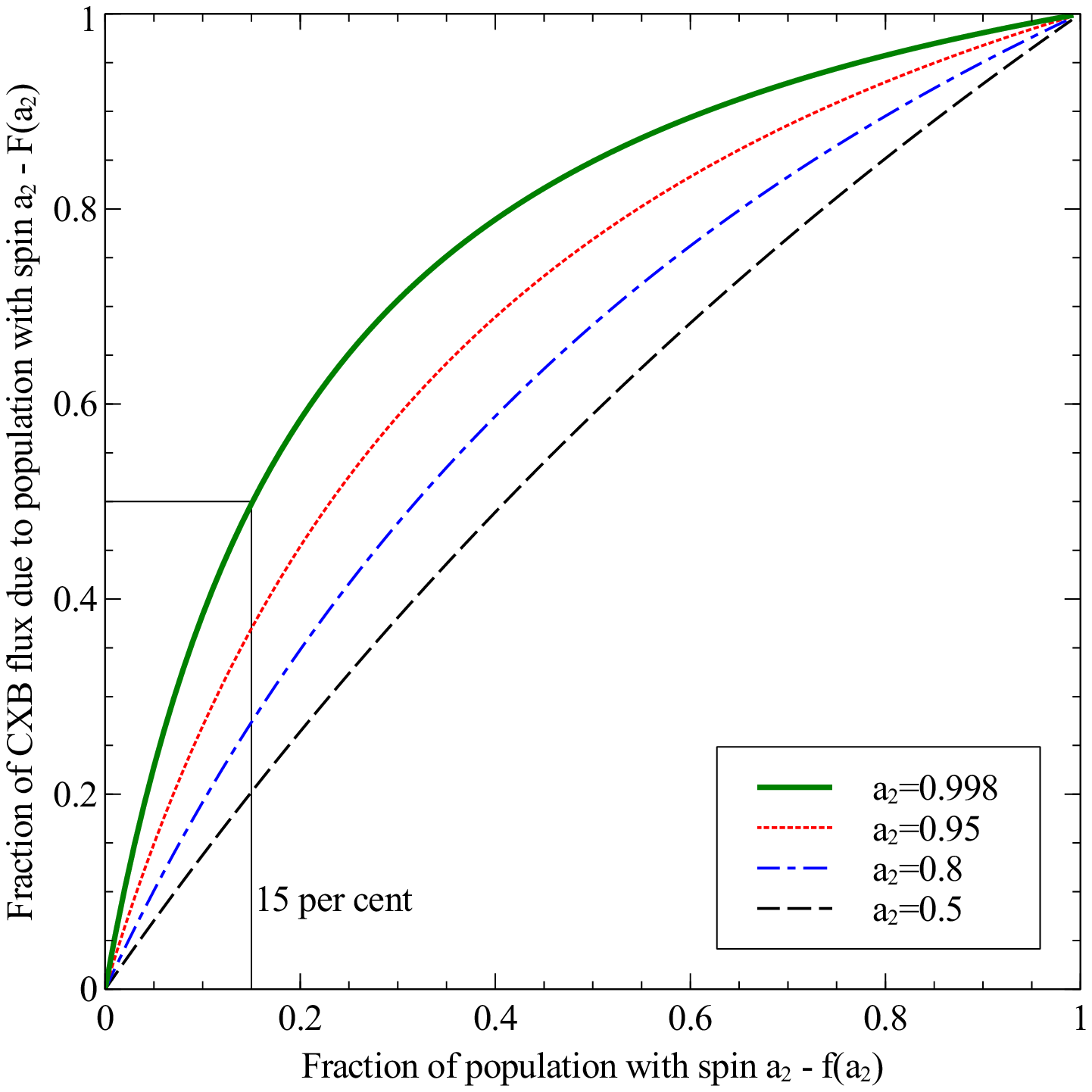}
}
\caption{Fraction of CXB intensity produced by high spin in a given redshift shell, as a function of the fraction of high spin objects.  The spin of the high spin population is varied to illustrate the effects of the change in radiative efficiency: dashed black line shows $a_{2}=0.5$ ($\eta=0.08$), dash-dotted blue line shows $a_{2}=0.8$ ($\eta=0.12$), dotted red line shows $a_{2}=0.95$ ($\eta=0.19$) and the solid thick green line shows $a_{2}=0.998$ ($\eta=0.32$, maximal spin).  The solid black line shows the required contribution of high spin to produce 50 per cent of the CXB energy density.\label{fig:FXRBvsfhighspin_norefl}}
\end{figure}

We finally calculate the dependence of the flux fraction due to high spin, $F(a_{2})$ as a function of the fraction of high spin objects $f(a_{2})$ (see Fig.~\ref{fig:FXRBvsfhighspin_norefl}), for different values of $a_{2}$.  The efficiency correction significantly boosts the contribution of high spin, such that a population of 15 per cent of maximally spinning black holes can make up 50 per cent of the CXB intensity.

\subsection{Band-specific effects of high spin in CXB synthesis modelling}
\label{bandeff}

In addition to high accretion efficiency boosting the total bolometric luminosity, the spectral shape in specific bands also changes.  The X-ray spectrum can change due to the availability of new accretion geometries allowed by high spin.  Specifically, as the innermost stable circular orbit (ISCO) can now move inwards to just over $\sim 1.24 R_{\rm g}$ for maximal spin, the X-ray source can also be located closer to the black hole. The emission from it can be subject to strong-gravity effects such as light-bending can, and the ionised reflection spectrum that can result from such effects will act to produce a host of band-specific spectral features.  We have already discussed how the shape of a strongly light-bent reflection spectrum will act to give a shape closer to that of the CXB spectrum itself, but the crucial question is whether an X-ray source close to the black hole will act to produce higher or lower flux in particular parts of the X-ray spectrum, as this will directly affect the contribution such sources make in those energy bands and their detectability in surveys taken in those bands.

For standard Shakura-Sunyaev accretion discs (without extreme super-Eddington or extremely low accretion rates) an underlying global constraint on the bolometric luminosity is
imposed by the accretion efficiency $\eta$, which is solely a function
of black hole spin.  
Therefore, in a global sense, the energy obtained from accretion onto a rapidly-spinning black hole with spin $a_{2}$ will be greater than that from
a spin-zero black hole ($a_{1}=0$) by the ratio of their efficiencies, $\eta_{2}/\eta_{1}$.  Whether this holds for the X-ray band is not immediately clear.
If the X-ray source is located closer to the black hole, it is possible that it will brighten due to increased exposure to the seed photons from the accretion disc, and additionally the reflection spectrum will brighten as the disc is reciprocally irradiated at closer quarters by the X-ray source.  Conversely, increased gravitational redshift and a greater proportion of light-paths leading into the black hole will also act to reduce the primary X-ray flux.  In the absence of detailed information on these effects, it is therefore reasonable as an ab-initio assumption to require that the total X-ray luminosity $L_{\rm X}$ obeys the global constraint: i.e., it is greater for a high-spin black hole than a low-spin black hole in proportion to the ratio of the accretion efficiencies, as holds for the \emph{total} luminosity $L_{\rm bol}$ from accretion:


\begin{equation} \label{eq:fluxratiohighlowspin}
\frac{L_{\rm X}(a_{\rm 2})}{L_{\rm X}(a_{\rm 1})} = \frac{L_{\rm bol}(a_{\rm 2})}{L_{\rm bol}(a_{\rm 1})} = \frac{\eta(a_{\rm 2})}{\eta(a_{\rm 1})}
\end{equation}

In order to understand the detailed implications on the X-ray spectrum, we make two key assumptions: 1) that a constant fraction of the power liberated by the accretion
disc appears in the corona (X-rays) irrespective of spin, and 2) that the X-ray source lies at a distance from
the black hole set primarily by the inner radius of the accretion disc, i.e., that the 
X-ray source lies $h \approx 2 \times R_{\rm isco}$ away from the black hole for any given spin (i.e., 12 $R_{\rm g}$ for zero spin and 2.5 $R_{\rm g}$ for maximal spin).  For the former assumption, we currently do not have any
constraints to the contrary so it is therefore the simplest assumption to make, and the second assumption is consistent with results from X-ray reverberation (see introduction).
We use the latest version of the {\sc relxilllp} model (v0.4a, \citealt{2014MNRAS.444L.100D,2016arXiv160103771D})
to calculate the spectral shape for spins $a_{\rm 2}$ and $a_{\rm 1}$.  In this calculation, and throughout this paper, we set the {\sc relxilllp} model parameter {\sc FixReflFrac} to be 1, which geometrically links the strength of the reflection features to the source height $h$.

In order to satisfy condition (\ref{eq:fluxratiohighlowspin}), we re-normalise the high spin ($a_{\rm 2}$) spectrum such that
it has a \emph{total} (0.1--500~keV) X-ray flux $\eta(a_{\rm 2})/\eta(a_{\rm 1})$ times more than the flux from the spin $a_{\rm 1}$
spectrum (hence the luminosities will be in the same ratio).  We then calculate the ratio of fluxes in specific bands, to investigate if they differ significantly from the overall $\eta(a_{\rm 2})/\eta(a_{\rm 1})$ ratio.   

\begin{figure}
\centerline{
\includegraphics[width=9.0cm]{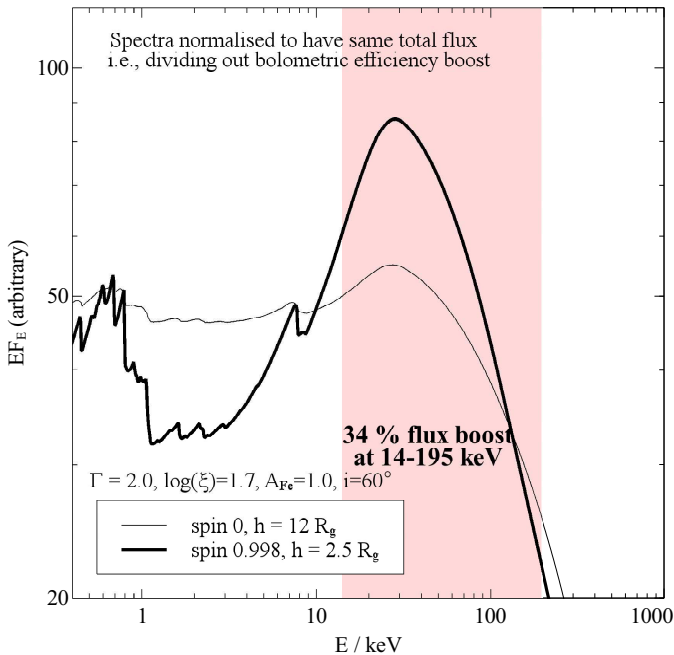}
}
\caption{X-ray spectra for zero and maximally spinning black holes, using the latest version of the {\sc relxilllp} model assuming an X-ray source at $h=2 R_{\rm isco}$ above the black hole (i.e. $12 R_{\rm g}$ for zero spin, and $2.5 R_{\rm g}$ for maximal spin), with $\rm log \xi = 1.7$, $i = 60^{\circ}$, $A_{\rm Fe}=1.0$ and $\Gamma = 2.0$.  The spectra have been normalised to have the same flux to illustrate the extra hard X-ray detectability boost.\label{fig:boostfactor}}
\end{figure}

For spins $a_{\rm 2}=0.998$ (maximal) and $a_{\rm 1}=0.0$, the overall flux boost factor should be 5.61 (i.e., the ratio of accretion efficiencies).   The boost in a specific band depends on the exact spectral shape, governed by the source height $h$, ionisation parameter $\rm log \xi$, inclination angle $i$, iron abundance $A_{\rm Fe}$ and photon index $\Gamma$.  For typical parameters $h = 3 R_{\rm isco}$, $\rm log \xi = 1.7$, $i = 60^{\circ}$ (e.g., see \citealt{2013MNRAS.428.2901W}), $A_{\rm Fe}=1.0$ and $\Gamma = 2.0$, we find that the BAT-band flux (14--195~keV) is boosted by a factor 7.5, which is 34 per cent greater than the bolometric boost (see Fig.~\ref{fig:boostfactor}). The 14--195~keV band can be boosted by up to a maximum of $\sim$ 60 per cent depending on the precise spectral parameters chosen.  For the same spectral parameters, the 2-10~keV band total boost is 4.65, which is 17 per cent lower than the bolometric boost of 5.61.  We also consider the effect of varying other spectral parameters on the 14--195~keV emission: ionisation parameters $\rm log \xi < 2.6$ enhance the overall efficiency boost whereas higher $\rm log \xi$ attenuates it; inclination angles $i>60^{\circ}$ can produce a boost enhancements of up to 58 per cent, and larger (softer) photon indices also increase the boost.  However, the fiducial parameters chosen to calculate the above numbers are typical of the lower-luminosity objects expected to make up the bulk of the CXB.  If one considers a narrow band of width 1~keV centred around 30~keV coincident with the peak of the reflection hump, the boost is 58~per~cent.

The 14--195~keV boost is naturally affected by the choice of higher spin $a_{2}$.  Adopting $a_{2}$ lower than 0.998 will reduce the boost effect: for comparison, if we choose a spin of $a=0.95$, the 14--195~keV band additional boost for the same fiducial parameters (assuming $h=2 \times R_{\rm ISCO}$) is 16 per cent, and for $a=0.9$, the boost is 12 per cent over the zero-spin case.

An outflowing corona \citep{1999ApJ...510L.123B} may dampen the extra hard X-ray  boost, as the reflection hump strength in such scenarios will be reduced compared to a `stationary' corona. The shape of the spectrum will be less peaked in such sources but their per-source luminosities will still be higher due to the overall efficiency correction.  However, outflowing coronae have not been widely reported in the objects for which spin has been measured so far, so such effects are unlikely to predominate.  Despite the recently reported detection of an outflowing corona in Mrk 335 in a flaring state \citep{2015MNRAS.454.4440W}, there is no evidence yet that this is typical/average behaviour for this object, or in other AGN.

The extra boost to the spectrum in hard X-rays will be typically less than a few tens of per cent, but when multiplied by the overall boost from $\eta$, it will cause additional CXB sensitivity to high spin.  AGN with rapidly-spinning black holes are therefore likely to be \emph{further} over-represented in the 14--195~keV CXB than the efficiency effect alone predicts.  Assuming $h = 2 R_{\rm isco}$, we calculate that 12 per cent can produce half of the 14--195~keV CXB, compared to 15 per cent without this effect.

\subsection{Implications for CXB synthesis models}
\label{implications_cxbsynth}

AGN luminosity functions (LFs) determined from measured fluxes, redshifts and absorption distributions can be used to synthesise the CXB spectrum, assuming a model for the X-ray spectra of the individual sources.  The spectral model is necessary for extrapolating from the number counts and fluxes in a specific (usually lower-energy) band to a broad-band spectral shape at higher energies, which is key for synthesising the broad-band CXB spectrum over its entire energy range.  

Current synthesis models use LFs determined at soft energies (below $\sim$8~keV) to reproduce the CXB spectrum by assuming a variety of absorbing column densities ($N_{\rm H}$) and reflection strengths (e.g., \citealt{2012A&A...546A..98A}).  Reproducing the peak of the CXB spectrum at $\sim$30~keV (significantly above the 0.5--8~keV energies where LFs are well constrained) is particularly challenging.  Existing CXB synthesis models utilise the strong resemblance between the shape of the strong peak in the CXB and that for Compton-thick AGN, and thus invoke a large population of Compton-thick AGN to fit the CXB peak (up to $\sim$40 per cent, e.g. \citealt{2007A&A...463...79G}).  The remainder of the CXB spectrum can be made up from populations with various absorptions and reflection strengths.

This work illustrates that high spin objects can provide an alternative population to explain the peak.  Fig.~\ref{relxill_cxb_comparison} illustrates how the spectral shape of high spin objects with pronounced reflection (due to a low coronal source height $h$) can directly contribute to the peak, and \cite{2007MNRAS.382.1005G} present a far more extensive treatment of how light-bending can produce a spectral \emph{shape} that can reproduce the CXB peak.  However, previous works did not consider the intrinsic brightness boost due to high spin, which we include in our treatment here.  When considering how strongly-peaked spectra such as light-bent or Compton-thick shapes can contribute to explaining the CXB peak, we emphasise that whilst the observed luminosity of Compton thick sources is comparable or less than those of less-absorbed sources, the observed luminosity of AGN with rapidly spinning black holes must (because of their efficiency) be significantly higher than that for AGN with non-rotating black holes, for comparable mass accretion rates.  Whilst Compton-thick objects can be invoked to match the peak of the CXB, gathering the required luminosity to produce the peak can require large numbers of such sources, which are not currently found in flux-limited hard X-ray surveys (e.g., \citealt{2011ApJ...728...58B}, Vasudevan et al. 2013a).  On the other hand, high-spin can be invoked to produce a shape that matches the CXB without requiring large numbers of such sources, due to their intrinsically high per-source luminosities.

We note that existing CXB synthesis models rely on the soft X-ray LFs and therefore a key component is the extrapolation of the spectrum to high energies based on an assumed spectrum and parameter distributions for the AGN populations; models have not yet been directly generated using LFs gathered at the peak of the X-ray emission at $\sim$30~keV, as current high-energy data do not probe sufficiently low fluxes to resolve the bulk of the population producing the CXB at these energies.  The \cite{2009ApJ...696..110T} and \cite{2012A&A...546A..98A} synthesis models show how one such appropriate parameterisation is in terms of varying absorption levels and phenomenological reflection strength $R$, most recently finding that reflection can be invoked as an alternative to large Compton-thick fractions.  Here we explicity point out that this reflection strength can be directly explained by light-bending around rapidly-spinning black holes, and that this explanation comes with a \emph{requirement} that the more rapidly spinning sources must contribute at an intrinsically higher luminosity (up to 5.6 times more than non-rotating black holes) at all redshifts where they exist.  In summary small populations of high spin AGN can be invoked where previously large numbers of Compton-thick sources or unexplained reflection strengths were required.  Undoubtedly, some combination of the two effects will ultimately be responsible for the CXB peak. Whilst existing synthesis models have yielded significant progress in our understanding, a CXB synthesis approach that includes the efficiency brightening effect is now necessary.

\subsubsection{A Toy CXB model with two spin populations}
\label{toycxbmodel}

\begin{figure*}
\centerline{
\includegraphics[width=9.0cm]{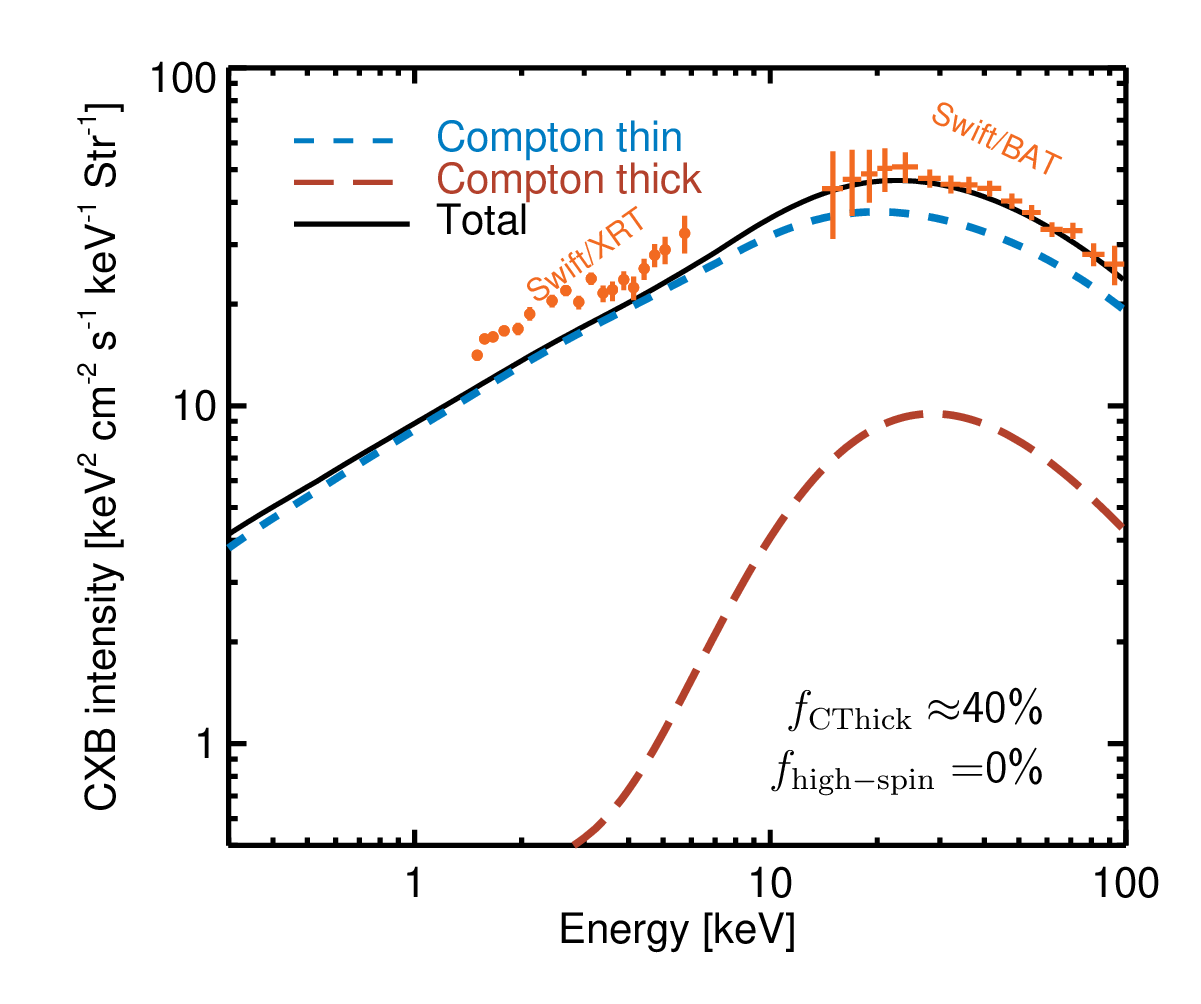}
\includegraphics[width=9.0cm]{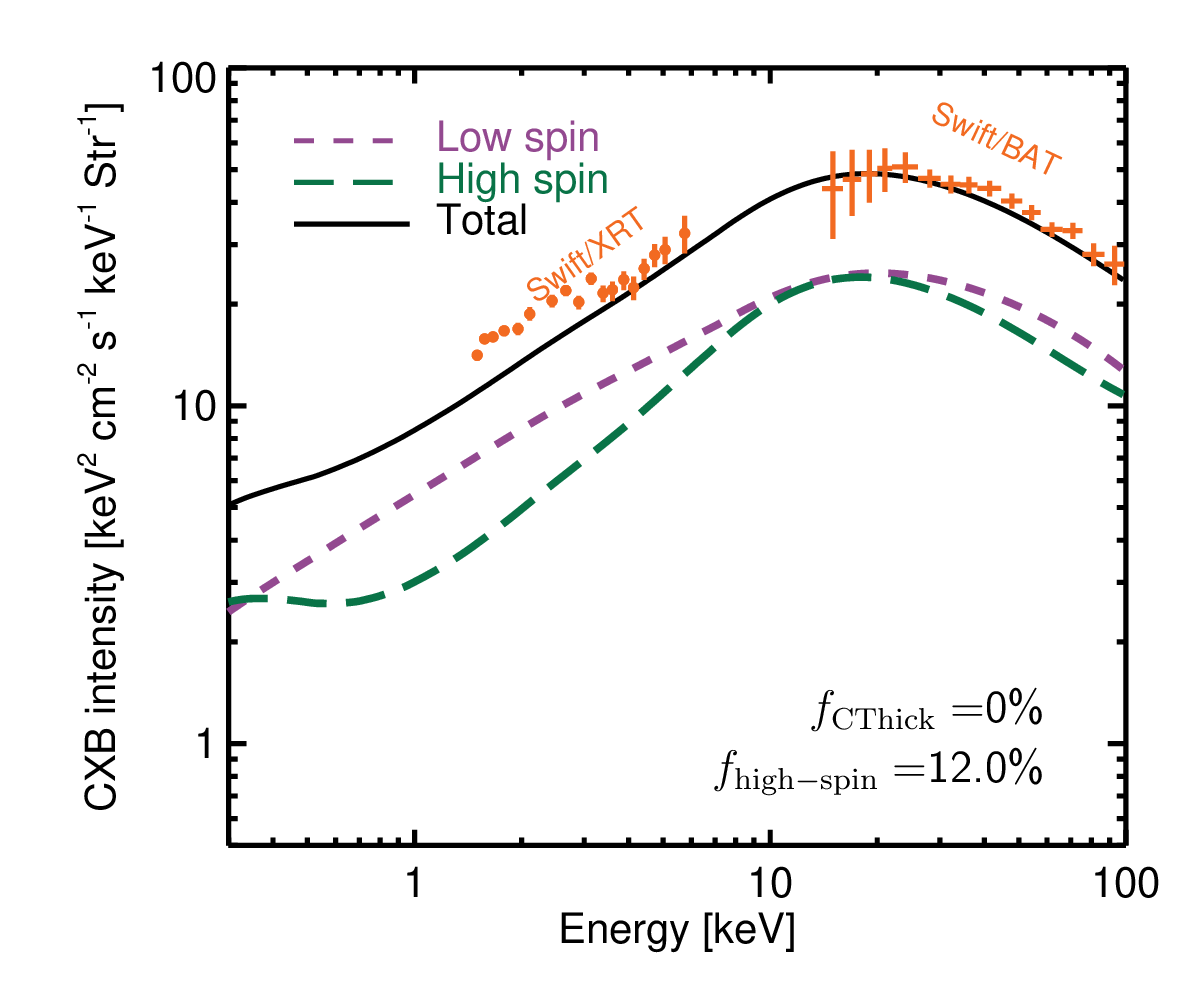}
}
\caption{\textit{Left:} The total predicted spectrum of the CXB using a population synthesis model on measurements of the X-ray luminosity function and the $\mathrm{N_H}$ distribution by \protect\cite{2014ApJ...786..104U}. Around 40\% of the AGN population is attributed to Compton-thick sources, which are needed to reproduce the peak of the CXB at $\sim20-30$~keV. \textit{Right:} Our toy model for the CXB, where we remove Compton-thick sources and attempt to reproduce the flux of the CXB using a combination of low- and high-spin sources, with the same underlying distribution of mass accretion rates. We are able to reproduce the CXB with $\sim12$\% of black holes having high spins, which nevertheless produce $\sim 40$\% of the flux of the CXB. Orange points in both panels show recent measurements of the total observed CXB from \emph{Swift}/XRT \citep{2009A&A...493..501M} and \emph{Swift}/BAT \protect\citep{2008ApJ...689..666A}.
\label{cxb_lf_highspin}}
\end{figure*}

To illustrate this potential impact of high spin objects on the CXB, we have developed a simple, toy model that allows for a contribution from a population of high-spin sources. We start with a standard population synthesis model based on the 2--10~keV X-ray LF and the distributition of $N_{\rm H}$ from \cite{2014ApJ...786..104U}, where around 40 per cent of the AGN population is associated with Compton-thick sources (depending on both luminosity and redshift). We adopt the X-ray spectral model for individual sources from \cite{2015MNRAS.451.1892A}, which allows us to reproduce the CXB spectrum over the $\sim$1--100~keV energy range (see Fig.~\ref{cxb_lf_highspin}, left panel).\footnote{We note that our model is slightly below the latest measurements of the background at soft X-ray energies ($<10$~keV) with Swift/XRT by \protect\cite{2009A&A...493..501M}, although is in good agreement with earlier measurements (e.g., \protect\citealt{1995PASJ...47L...5G}) that had a lower normalisation over this energy range.} Fig.~\ref{cxb_lf_highspin} (right panel) shows our toy model where we instead split the AGN population into low ($a=0$) and high ($a=0.998$) spin sources. We use the \cite{2014ApJ...786..104U} LF along with a single efficiency (for spin $a=0$) to estimate the shape of the mass accretion rate ($\dot{M}$) distribution.  This $\dot{M}$ distribution is then scaled by two efficiencies for maximal and zero spin sources, to produce a `synthetic' LF that now includes both spin populations.  This therefore assumes that the (intrinsically) small fraction of maximally-spinning sources has the same $\dot{M}$ distribution as all other sources, but we require that the \emph{observed} 2--10~keV luminosities are boosted at a given $\dot{M}$ compared to the low-spin sources by a factor 4.65 (accounting for the overall difference in efficiencies ($\eta$) with a small additional correction for the 2--10~keV band, see \S\ref{bandeff}). The {\sc relxilllp} X-ray spectral model is adopted for the high-spin sources, assuming $h=2.5 R_{\rm g}$ (2 $\times$ ISCO), with other parameters as in \S\ref{bandeff}. We assume the same distribution of $\mathrm{N_H}$ for both zero- and maximal-spin populations but remove any Compton-thick ($\mathrm{N_H}>10^{24}$ cm$^{-2}$) sources. Finally, we scale the relative, intrinsic fractions of low- and high-spin sources (at a given $\dot{M}$) such that our toy model reproduces the same CXB flux at 2~keV and 20~keV as in the original model (shown in Fig.~\ref{cxb_lf_highspin} left). With this toy model, we can recover the CXB spectrum with $\sim12$ per cent of black holes having high spin. Nonetheless, this relatively small high-spin population produces $\sim$45 per cent of the total 1--200~keV flux, consistent with our estimates from Fig.~\ref{fig:FXRBvsfhighspin_norefl}.

We note that our toy model is intended to demonstrate the most extreme possibility, rather than being a fully accurate and self-consistent model. In reality, we would expect a distribution of spins spanning the full range (although as discussed above, high-spin objects are likely to be over-represented in observed samples). Furthermore, we know that Compton-thick AGN do exist and must contribute to the CXB at some level, although the exact fraction of the AGN population is still unclear (especially at high redshifts, see e.g., \citealt{2009ApJ...696..110T}, \citealt{2015ApJ...802...89B}, \citealt{2015MNRAS.451.1892A}). The key point of our toy model is to demonstrate that an \emph{intrinsically} small fraction of high-spin objects can produce the characteristic “hump” of the CXB, where a relatively high fraction of Compton-thick AGN would usually be required. It is vital that future studies carefully consider the underlying distribution of spin---in addition to the absorption distribution---when constructing population synthesis models that fully satisfy the constraints from the observed LF and CXB.

\section{High spin in flux-limited surveys}
\label{fluxlimsurv}

\cite{2011ApJ...736..103B} discuss the bias towards high spin
introduced to flux-limited X-ray surveys by the inclusion of the
efficiency correction.  A flat (uniform) spin distribution between
zero and maximal spin produces an observed spin
distribution with half of the objects having spins greater than 0.73,
even though in the parent population half will have spins greater than
$0.5$.  If a spin distribution $f(a) \propto a$ is employed, half of
the observed objects will have spins above 0.84, even though in the
parent population half will have spins above 0.70.

We apply their approach to the simple two-spin population considered
in \S\ref{effcorr} to identify the degree of over-representation of
high spin in flux-limited surveys.  For a Euclidean universe (valid
for local/bright AGN samples in which spin can reasonably be
determined using current facilities), \cite{2011ApJ...736..103B}
identify that the number of \emph{detected} AGN in a flux-limited survey
in the spin range [$a$,$a+da$] is:

\begin{equation} \label{eq:fluxlimsurveyN}
dN_{a} \propto f(a) \eta(a)^{3/2} da
\end{equation}
where $f(a)$ is the flux distribution, i.e. $f(a)da$ is the number of
sources in the population with spins in the same range [$a$,$a+da$].
The factor $\eta^{3/2}$ arises from the assumption that the
distribution of sources is uniform in Euclidean space. It breaks down
when the counts extend below a flux level corresponding to the most
distant high luminosity source. If we apply this to a simple two-spin
distribution, we can identify that the number of sources with spin
$a_{i}$ is:

\begin{equation} \label{eq:fluxlimsurv_generalN}
N(a_{i}) \propto f(a_{i}) \eta_{i}^{3/2} 
\end{equation}
where $\eta_{i}$ is the radiative efficiency corresponding to spin $a_{i}$, and $i=1,2$ to represent our two fiducial spins.  We can straightforwardly determine the fraction of sources with high spin (taken to be $a_{2}$) using the approach in \S\ref{effcorr} and find that:

\begin{equation} \label{eq:fluxlimsurvey_finalformula}
F(a_{2}) = \frac{f(a_{2}) \eta_{2}^{3/2}}{(1-f(a_{2})) \eta_{1}^{3/2} + f(a_{2}) \eta_{2}^{3/2}}.
\end{equation}

\begin{figure}
\centerline{
\includegraphics[width=8.0cm]{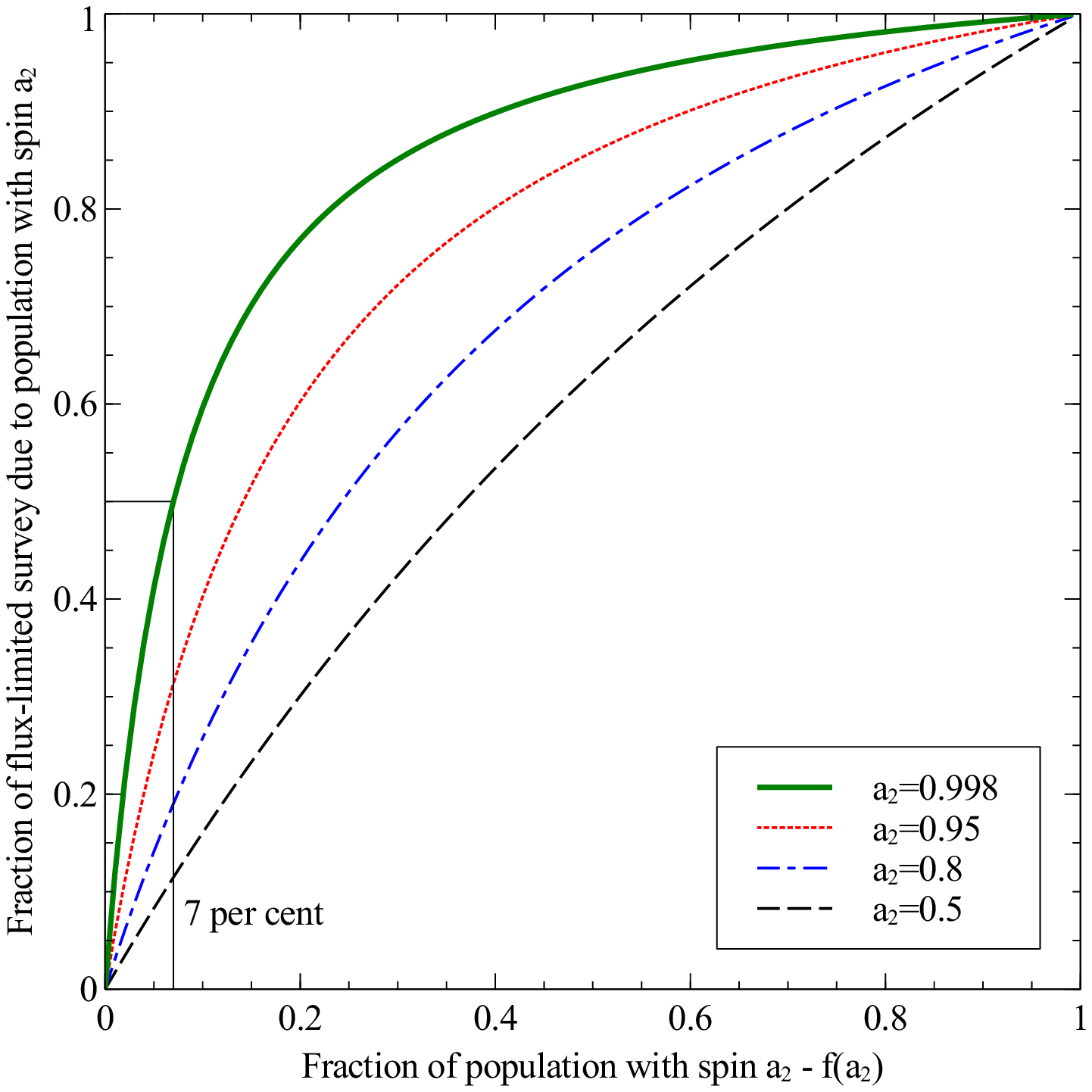}
}
\caption{Fraction of a flux-limited survey produced by high spin, as a function of the fraction of high spin objects.  Key as in Fig.~\ref{fig:FXRBvsfhighspin_norefl}.\label{fig:fluxlimsurveyhighspinfrac}}
\end{figure}

Again considering $a_{1}=0.0$ and $a_{2}=$0.5, 0.8, 0.9 and 0.998, we can produce the expected fraction of sources with high spin in a flux-limited survey (Fig.~\ref{fig:fluxlimsurveyhighspinfrac}).  Because of the $\eta^{3/2}$ factor, the over-representation of high spin is even more pronounced: 7 per cent of maximally-spinning black holes in a population with the remainder having zero spin, will make up 50 per cent of the number counts in a flux-limited survey.

\subsection{Band specific effects in flux-limited surveys}
\label{bandeff_surveys}


The band-specific effects highlighted in \S\ref{bandeff} with reference to the CXB will have a directly analogous effect on the number counts obtained from surveys in different bands, but the effect will be even more pronounced, as the efficiency correction (including any band-specific variation) enters into the selection function raised to the power 3/2. Again assuming $h = 2 R_{\rm isco}$, we calculate that 4.7 per cent maximally-spinning black holes can produce half of the 14--195~keV survey number counts respectively, compared to 7 per cent without this additional boost.




\subsection{The importance of effiency for luminosity functions, black hole mass functions and the CXB}
\label{luminosityfunctions}

AGN LFs record the number counts of AGN as a function of luminosity and redshift, and therefore provide a measure of the evolution of accretion output from supermassive black holes.  Black hole mass functions (MFs) measure the number counts of black holes as a function of black hole mass and redshift, and therefore if one assumes that the accretion output recorded by the LF is responsible for the observed black hole mass buildup in the MF, one can match up the two (e.g., \citealt{2008MNRAS.388.1011M}) to say something about the accretion process responsible for producing the black hole mass buildup.  This has been done by many authors, generally assuming that a single efficiency can be applied to the whole population; for example \cite{2002ApJ...565L..75E} found that the average accretion efficiency must be $\eta_{\rm avg}>0.15$, requiring spins above 0.88 on average.  \cite{2004MNRAS.351..169M} later find that the LF can match the MF for efficiencies $\eta = 0.04-0.16$, without requiring rapidly-spinning black holes, but with spinning black holes nevertheless producing the best fit between the relic black hole MF and the X-ray background and LF, on average.

Whilst adopting a single accretion effiency has previously produced a successful match between LFs and MFs, if there is a wide spin distribution in the real AGN population, this assumption will need to be relaxed to obtain more physically meaningful results.  A single efficiency with a spread assumes that all AGN in the LF produce the same amount of flux for a given mass accretion rate, and once multiple spin populations are invoked, this assumption breaks down.  The existing, single best-fit efficiencies obtained from LF/MF matching will present an average picture, but will not be able to incorporate the highly non-linear way in which the luminous output depends on spin. As highlighted in this paper, it is the \emph{ratio} of efficiencies between different spin populations which makes the high-spin bias evident, and a single radiative efficiency offers no scope for different spin populations to be distinguished, if present.  \cite{2009MNRAS.396.1217R} show that the introduction of even two spin populations in LF-MF calculations introduces substantial degeneracies when trying to reproduce the observed local black hole mass distribution, but find that assuming a fixed $\eta$ or Eddington ratio is insufficient for properly characterising BH evolution, nor is it consistent with the observation of low-Eddington-ratio objects.  Despite these difficulties, it is clear that the effects of over-representation of even a small high spin population are inescapable, and future models will require a careful (and probably degenerate) balancing of spin and absorption to satisfy all the available observational constraints.

Previous conclusions on mass build-up from CXB-MF/LF matching (\citealt{1982MNRAS.200..115S}, \citealt{2002ApJ...565L..75E}, \citealt{2004MNRAS.354.1020S}, \citealt{2004MNRAS.351..169M}) may need significant revision when multiple populations with different spins and radiative efficiences are considered.  If a smaller number of high-spin sources can now be invoked to reproduce the CXB peak, we know that these sources must be responsible for less mass build up: the higher radiative efficiency due to rapid spin means that less mass is built up in rapidly-spinning sources.  Therefore the rapidly-spinning sources that are preferentially represented in the CXB spectrum and in LFs would not be sufficient in themselves to match the observed mass build-up in the local black hole mass function. This shortfall in the black hole mass density could then require a substantial population of `quiet', low-spin accretion such as ultra-Compton-thick AGN that would be missed in both surveys and the CXB (e.g., \citealt{2015A&A...574L..10C}), although this would be a difficult hypothesis to test by observation.  

Disentangling these effects is non-trivial. As shown in \S\ref{fluxlimsurv}, the number counts in flux-limited surveys are heavily affected by the spin bias, and therefore existing LFs will be contaminated by efficiency bias, especially hard X-ray selected surveys above 10~keV as discussed in \S\ref{bandeff}. In order to correct for this bias in matching LFs and MFs, one needs to assume a spin distribution and assign each object with its own spin/radiative efficiency, rather than adopting a single average spin for the whole population as has been commonly done previously. Whilst we have explored these effects briefly in our toy CXB synthesis in \S\ref{toycxbmodel}, fully incorporating these extra degrees of freedom into the calculations is beyond the scope of this work.


\section{Indications of high spin in samples of AGN from X-ray reflection}
\label{highspinsamples}



\begin{figure}
\centerline{
\includegraphics[width=0.5\textwidth]{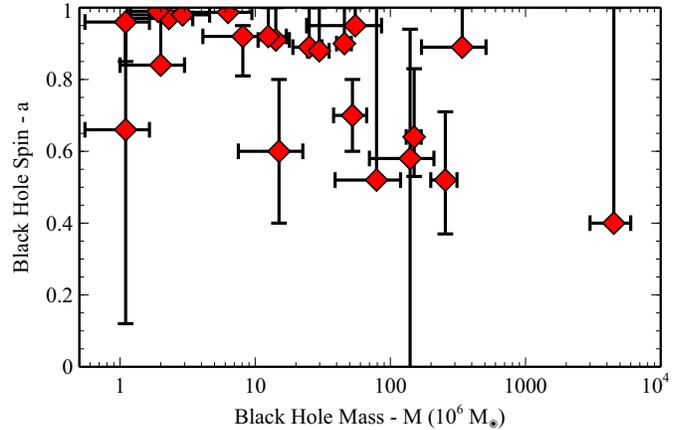}
}
\caption{Plot of SMBH mass $M$ and spin $a$ from the sample listed in Table~\ref{tab:1} (the three objects without mass estimates are omitted from the figure).   Reflecting the conventions in the primary literature, all masses are marked with $1\sigma$ error bars whereas spins are marked with $90$ per cent error ranges.   When no error estimate is available for the mass, we have assumed an error of $\pm 0.5M$.  }
\label{fig:smbh_spins}    
\end{figure}

Detailed, high signal-to-noise spectroscopy of about two dozen
individual local AGN in the Fe K$\alpha$ band has revealed that most
of them host rapidly spinning black holes; \cite{2013CQGra..30x4004R}
summarise the current spin measurements available from detailed
spectroscopy of the Fe K$\alpha$ line region in individual AGN.  We
now have robust constraints on spin in 25 moderately-luminous AGN,
finding high spin ($a>0.9$) in many (12) of the objects.  Table~\ref{tab:1}
summarises the latest spin measurements from Fe K$\alpha$ line
analysis including more recent spin determinations not available to \cite{2013CQGra..30x4004R}, and
Fig.~\ref{fig:smbh_spins} shows the spins as a function of black hole
mass.  At higher masses, there may be some intermediate
spins. However, spin measurements from this method rely on very
high-quality data to measure the red wing of the Fe K$\alpha$ line,
restricting this approach to mainly bright objects.

\begin{table*}
\begin{center}{
\caption{Summary of published AGN/SMBH spin measurements.  All measurements are based upon {\it XMM-Newton} and/or {\it Suzaku} data.  Reflecting the conventions in the primary literature, all masses are quoted with $1\sigma$ error bars whereas spins are quoted with 90 per cent error ranges. Key to references: AG14=\protect\cite{2014MNRAS.443.2862A}; Be11=\protect\cite{2011ApJ...726...59B}; BR06=\protect\cite{2006ApJ...652.1028B}; Br11=\protect\cite{2011ApJ...736..103B}; Fa13=\protect\cite{2013MNRAS.429.2917F}; Ga11=\protect\cite{2011MNRAS.411..607G}; Go12=\protect\cite{2012A&A...544A..80G}; Lo12=\protect\cite{2012ApJ...758...67L}; Lo13=\protect\cite{2013ApJ...772...83L}; Ma08=\protect\cite{2008MNRAS.389.1360M}; Mc05=\protect\cite{2005MNRAS.359.1469M}; Mi09=\protect\cite{2009MNRAS.398..255M}; Pe04=\protect\cite{2004ApJ...613..682P}; Pa12=\protect\cite{2012MNRAS.426.2522P}; Re14=\protect\cite{2014ApJ...792L..41R}; Ri09=\protect\cite{2009ApJ...696..160R}; Ri13=\protect\cite{2013Natur.494..449R}; Ri14=\protect\cite{2014ApJ...795..147R}; Wa13=\protect\cite{2013MNRAS.428.2901W}; Zo10=\protect\cite{2010MNRAS.401.2419Z}; ZW05=\protect\cite{2005ApJ...618L..83Z}; Be06=\protect\cite{2006ApJ...651..775B}; Ke15=\protect\cite{2015arXiv150407950K}.\label{tab:1}}
}
\begin{tabular}{lcccl}
\hline\noalign{\smallskip}
Object  & Mass ($\times 10^6M_\odot$) & Spin & Mass/Spin References  \\
\noalign{\smallskip}\hline\noalign{\smallskip}
Mrk335  		& $14.2\pm 3.7$	& $>0.91$*	& Pe04/Ga15 \\
IRAS~00521--7054 & ---			& $>0.73$				& --/Ri14\\
Tons180 		& $\sim 8.1$		& $0.92^{+0.03}_{-0.11}$	& ZW05/Wa13 \\
Fairall~9		& $255\pm 56$		& $0.52^{+0.19}_{-0.15}$**	& Pe04/Lo12 \\
Mrk359		& $\sim 1.1$ 		& $0.66^{+0.30}_{-0.54}$	& ZW05/Wa13 \\
Mrk1018		& $\sim 140$		& $0.58^{+0.36}_{-0.74}$	& Be11/Wa13 \\
NGC1365		& $\sim 2$		& $>0.84$				& Ri09/Ri13 \\
1H0419-577 	& $\sim 340$		& $>0.89$				& ZW05/Wa13\\
3C120		& $55^{+31}_{-22}$	& $>0.95$				& Pe04/Lo13 \\
Ark120		& $150\pm 19$		& $0.64^{+0.19}_{-0.11}$	& Pe04/Wa13\\
Swift~J0501.9-3239 & ---			& $>0.99$				& --/Wa13 \\
1H0707-495	& $\sim 2.3$		& $>0.97$				& ZW05/Zo10\\
Mrk79		& $52.4\pm 14.4$	& $0.7\pm 0.1$			& Pe04/Ga11\\
Mrk110		& $25.1\pm 6.1$	& $>0.89$				& Pe04/Wa13\\
NGC3783 	& $29.8\pm 5.4$ 	& $>0.88^*$ 			& Pe04/Br11 \\
NGC4051		& $1.91\pm0.78$	& $>0.99$				& Pe04/Pa12 \\
RBS1124		& ---				& $>0.97$				& --/Wa13\\
IRAS13224--3809 & $\sim 6.3$	& $>0.987$			& Go12/Fa13\\
MCG--6-30-15	& $2.9^{+1.8}_{-1.6}$& $>0.98$			& Mc05/BR06\\
Mrk841		& $\sim 79$		& $>0.52$				& ZW05/Wa13\\
Swift~J2127.4+5654 & $\sim 1.5$	& $0.6\pm 0.2$			& Ma08/Mi09\\
Ark564 		& $\sim 1.1$		& $0.96^{+0.01}_{-0.11}$	& ZW05/Wa13 \\
ESO~362--G18 & $12.5\pm 4.5$     & $>0.92$				& AG14/AG14 \\
H1821$+$643 & $4500\pm 1500$   & $>0.4$              		& Re14/Re14\\
NGC 4151  & $45.7^{+5.7}_{-4.7}$ & $>0.9$  & Be06/Ke15 \\
\noalign{\smallskip}\hline
\end{tabular}
\end{center}
* result for Mrk 335 consistent with more recent result
$a>0.98\pm0.01$ from \cite{2014MNRAS.443.1723P}.\\
** 2 solutions were found for Fairall 9; we tabulate the lower spin value here.
\end{table*}

\begin{figure}
\centerline{
\includegraphics[width=0.5\textwidth]{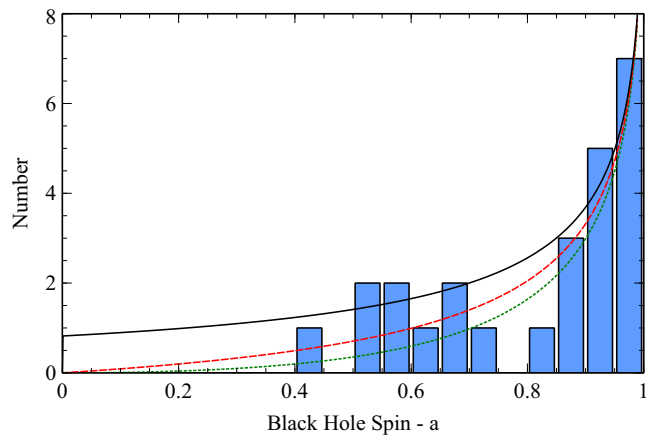}
}
\caption{Distribution of spin measurements from Table~\ref{tab:1}. The lines
  represent the observed spin distributions expected from an intrinsic distrubtion $f(a)\propto a^p$, with
  $p=0$ (black solid), 1 (red dashed) and 2 (green dotted). }
\label{fig:spin_histo}    
\end{figure}

We plot the distribution of spins in Fig.~\ref{fig:spin_histo} together with lines
representing the expected \emph{observed} distribution for a flux-limited sample with an \emph{intrinsic} spin distribution $f(a)\propto a^p$ with $p=0, 1$ and 2. In
principle the distribution extends down to $a=-1$. Several of the
sources in the range between $a=0.4-0.7$ have upper limits compatible with
much higher spin. The results so far therefore appear compatible with
a flat intrinsic distribution of spin from 0.4 to 1, or a power-law distribution
steeper than $f(a)\propto a$. The observed spin distribution instantly rules out
the possibility that all AGN have a single spin, and furthermore requires that there should be a substantial fraction
of unobserved lower-spin AGN that have been missed out due to the radiative efficiency bias.

 We note however that while the sources in
Table~\ref{tab:1} are among the brightest in the sky they do not consitute a
complete sample. Robust measurement of spin requires good irradiation of the
ISCO which in turn requires at least part of the corona to lie at $h<10 R_{\rm g}$ \citep{2014MNRAS.439.2307F}.  Since the smallest source heights are only accessible with high spin, this itself provides a
weak bias against low values of $a$, which may explain the absence of objects with $a \lesssim 0.5$ in Table~\ref{tab:1}.

In addition to these direct measurements, there are a host of observational results from AGN spectral stacking and sample analyses, that are consistent with the idea of an over-representation of high spin.  These are listed in detail in Appendix~\ref{app1}.

\section{Discussion}
\label{othereff}

We have thus far not considered any link between the mass accretion
rate $\dot{M}$ and spin itself, but it is conceivable that spin and $\dot{M}$ are linked through feedback or the mode of fuelling in some manner.  Additionally, a corona that is
co-rotating with the disc would have parts of it rotating at up to
$\sim 0.5c$, resulting in light being `thrown out' from the corona,
boosting the power-law component relative to the reflected emission.
The polar diagram of the emission from such a corona would be beamed
preferentially along the plane of the accretion disc, as seen in Fig.~10 of \cite{1997MNRAS.288L..11D}.  A simple solid
angle argument suggests that high-inclination sources would be seen
more, indicating that such boosted emission should be frequently seen.

It is interesting to note that the selection effects for spin in
supermassive black holes do not directly apply to Galactic (stellar
mass) black hole systems, for which different selection functions
apply.  Most such systems are transient, and their variability generally far exceeds the factor of 5 achievable from efficiency changes. This accounts for the wider range of spins observed for Galactic black hole candidates.

The efficiency boost considered here does not include any
consideration of the \cite{1977MNRAS.179..433B} process whereby
further rotational energy can be extracted from the black hole if the
magnetic field and spin are large enough. This may also occur if some
of the disc angular momentum passes into the corona and beyond rather
than all being transferred outward through the disc.  Inclusion of
these effects can produce even higher radiative efficiencies and
therefore provide an even larger efficiency boost than considered
here. We note that several models for powering radio jets have a strong spin dependence (see e.g., \citealt{2010ApJ...711...50T,2011MNRAS.418L..79T}, \citealt{2013ApJ...762..104S}).  This means that the selection effects discussed here are relevant to radio observations of jetted emission as well, if the models are correct.  

The effects discussed here are also relevant to the unobservable UV background due to quasars. \cite{2015arXiv150200637S} find that a UV Background enhanced by a factor of about two over the value calculated in \cite{2012ApJ...746..125H} is required to explain the observed distribution of Ly$\alpha$ absorbers. Some of this discrepancy could be accounted for by invoking a contribution from rapidly spinning black holes in the quasar population, since for such black holes, the far-UV excess will be boosted due to energy release from hotter regions of the accretion flow closer in to the black hole.

In the wider context of galaxy evolution, \cite{2014ApJ...794..104S} point out how taking the spin distribution of SMBHs into account has a direct effect on establishing the relative roles of black hole spin-up due to accretion vs. mergers.  However, they employ the spin distribution of 
\cite{2014SSRv..183..277R} as representative of the true (intrinsic) spin distribution; their conclusions may be therefore significantly influenced (and indeed enhanced) by the bias pointed out here.

\section{Summary}
\label{summary}

We emphasise that in both the CXB and flux-limited surveys, we are
subject to a strong selection effect towards high spin due to the
accretion efficiency. 
The increased radiative efficiency around rapidly-spinning black holes
should naturally over-represent them in the Cosmic X-ray Background,
such that a population of only 15 per cent maximally-spinning black
holes can produce 50 per cent of the CXB flux in any given redshift
shell.  In a flux-limited survey, a population of 7 per cent
maximally-spinning black holes can account for half of the sample.  
Assuming that the X-ray source is located $\sim 2 R_{\rm isco}$ above the black hole (as consistent with recent X-ray reverberation results) the hard X-ray ($>$10~keV) spectrum can be boosted further by $\sim$30 per cent, and up to $\sim$ 60 per cent (depending on other reflection model parameters) more than the overall efficiency boost, over-representing high spin even more in AGN surveys such as the \emph{BAT} or \emph{NuSTAR} catalogues. 

Recent CXB synthesis models that allow room for large values of
reflection (and lower Compton-thick fractions) are entirely consistent
with this finding.  We illustrate in a toy model that an intrinsically small fraction of maximally-spinning black holes can reproduce the CXB peak without invoking large
numbers of Compton-thick AGN.
The 25 nearby, bright objects for
which direct, Fe K$\alpha$ modelling can be done
\citep{2013CQGra..30x4004R} and Fig.~\ref{fig:spin_histo} reveal high spin in a majority of sources
as well.  A large body of observations (see Appendix~\ref{app1}) is consistent with
this effect. 

We reiterate that in order to measure spin accurately, we require small source heights $h \sim 2-10 R_{\rm g}$.  Since such heights are more available for high spin, this itself introduces a bias against spins below $\sim$0.5. These are all consistent with the idea that an intrinsically
small proportion of high-spin objects will be over-represented in
typical AGN samples.

The effects outlined in this paper have wide-ranging applicability and relevance, for CXB synthesis, AGN LFs, and even for understanding galaxy evolution. Since the LFs are derived from survey source counts, they too are subject to a strong spin-selection effect.  Fully accounting for the effects of spin in all these scenarios requires assumptions about the \emph{intrinsic} spin distribution which is currently unconstrained.  To make progress, we firstly require more robust constraints on the \emph{observed} spin distribution in order to shed light on what the intrinsic distribution is. This will be achieved by large flux-limited AGN samples with high-quality wideband X-ray spectra from \emph{NuSTAR} \citep{2013ApJ...770..103H} and \emph{ASTRO-H} \citep{2014SPIE.9144E..25T}.

\section{Acknowledgements}
We thank two anonymous referees for helpful comments and suggestions that improved this work.  RVV thanks the STFC for support. ACF and JA acknowledge ERC Advanced Grant FEEDBACK. CSR is grateful for financial support from the Simons Foundation (through a Simons Fellowship in Theoretical Physics), a Sackler Fellowship (hosted by the Institute of Astronomy, Cambridge), and NASA under grant NNX14AF86G.  CSR thanks the Institute of Astronomy at the University of Cambridge for their hospitality during an extended visit while this work was being performed. We thank Dom Walton for useful discussions on this work.

\appendix
\section{Observations consistent with preferential selection of high spin}
\label{app1}
\subsection{Stacked local broad-band AGN spectra}

Recent stacking analyses of hard X-ray selected AGN samples (\citealt{2013ApJ...770L..37V}, \citealt{2011A&A...532A.102R}) and X-ray background synthesis models (\citealt{2012A&A...546A..98A}, \citealt{2009ApJ...696..110T}) suggest that significant reflection can successfully fit the CXB spectrum.  We note that the selection function should be intrinsically different for the CXB and flux-limited samples: contribution to the CXB is proportional to $\eta$ whereas for flux limited samples the number counts depend on $\eta^{3/2}$ (see \S\ref{effcorr} and \S\ref{fluxlimsurv} above), suggesting that it is not obvious that these stacked spectra should look similar to the CXB. Nevertheless, these hard X-ray ($>$10~keV) catalogues are the most representative surveys of AGN since they are immune to the effects of host-galaxy dilution and absorption, which attenuate the fluxes at other bands.    High reflection, under the paradigm outlined in this paper, can be produced by light-bending around a rapidly spinning (radiatively efficient) black hole.  We first consider the unobscured (column density of neutral hydrogen $\rm log \thinspace N_{\rm H} \thinspace < 22$) stacked AGN spectrum from the BAT 58-month Northern Galactic Cap sample as representative of `clean' AGN, for which the reflection hump should be most clearly visible.  A fit to this spectrum with a high-spin reflection model ($a=0.9$) using a cut-off power-law model added to a relativistically blurred ioinised reflection component ({\sc cutoffpl + kerrconv(reflionx)} in {\sc xspec}) yields a good fit, showing that the BAT AGN are consistent with the survey preferentially sampling high spin due to increased radiative efficiency.

Interestingly, the \cite{2011A&A...532A.102R} and \cite{2013ApJ...770L..37V} analyses reveal a trend for the Compton hump strength $R$ to increase with absorption $N_{\rm H}$.  Conventionally, this has been interpreted as due to multiple reflections  from distant clumpy material in a circumnuclear dusty torus.  Assuming the `Unified Model' geometry where differences in absorption arise from different viewing angles, the conventional understanding of the increase in Compton hump strength with $N_{\rm H}$ is that for low inclination angles (i.e., close to face-on to the accretion disc), the reflection that is observed is due to a combination of ionised (disc) and neutral (torus) reflection  (at the level of $R \sim 1$ for unabsorbed objects).  As the incination angle increases, the absorption increases as the number of clumpy clouds intercepted by the line of sight increases.  In conjunction with this, there are many more pathways for multiple reflections from many clouds in a clumpy torus, which would act to boost the reflection fraction to the values of $R \sim 2$ seen for $10^{23} < N_{\rm H} < 10^{24} \rm cm^{-2}$.  The hard--X-ray peak of the CXB may also be consistent with multiple complex absorbers without invoking reflection, as described in \cite{2013ApJ...762...80T}, but we leave exploration of that paradigm to a future work.

Light-bending close to the black hole may also be invoked to explain this trend: for a corona at $\sim 3 R_{\rm g}$, relativistic beaming of the hard X-ray emission along the plane of the accretion disc (\citealt{1986ESASP.263..641S}) can cause more prominent hard excesses when the disc is viewed close to edge-on (see the discussion in \S\ref{bandeff} of this paper on the inclination angle dependence of the high-spin boost in the hard X-ray band).  Since edge-on angles correspond to higher absorption levels, this could explain the observed evolution of $R$ with $N_{\rm H}$.  To test whether the data is consistent with this, we again fit the cut-off power-law and ionised reflection model to the `medium obscured' class of AGN (`MOB' in the \citealt{2011A&A...532A.102R} and \citealt{2013ApJ...770L..37V} nomenclature) but this time including a partial covering absorber to model the low-energy fall-off. We again find that high spin is consistent with the spectral shape.  This is significant because $R\sim 1$ for unabsorbed AGN does not rule out neutral reflection, whereas $R \sim 2$ seen for MOB AGN would necessarily require light-bending, if due to inner-disc reflection.  That a successful fit is possible to the MOB AGN spectrum means that high spin is consistent with the large hard excess seen.  We emphasise that this does not rule out a low-spin fit to these stacked spectra, but we point out that they are nevertheless consistent with high spin.

\subsection{Stacked distant AGN spectra}

Stacked 2--10~keV spectra of AGN from various catalogues have all revealed strong evidence for a broad Fe K$\alpha$ component in addition to a narrow, neutral Fe K$\alpha$ emission line (e.g., \citealt{2005ApJ...621L...5B}; \citealt{2005A&A...432..395S}; \citealt{2012A&A...537A...6C}; \citealt{2012A&A...537A..86I}; \citealt{2013A&A...555A..79F,2014A&A...568A..15F}).  The broad component can be fit with relativistically blurred reflection originating in the accretion disc close to the SMBH (e.g. \citealt{2005MNRAS.358..211R}) and is expected to peak at 20~keV.    \cite{2012A&A...537A...6C} modelled the broad component in the 2XMM summed spectrum with a disc line around a Schwarzschild black hole, but found that both zero spin and maximal spin were equally good fits to the data.  That high spin is allowed by the data is entirely consistent with a scenario in which an intrinsically smaller population of rapidly-spinning black holes would be heavily over-represented due to their greater radiative efficiency.

The recent study of \cite{2015arXiv150305255W} shows that stacked \emph{Chandra} observations of lensed quasars in the redshift $1.0 \lesssim z \lesssim 4.5$ exhibit a pronounced red wing around 6.4~keV, consistent with ionised reflection from the inner parts of the accretion disc.  This would indicate that even at higher redshifts, there are potential indicators of high spin and light-bending which can be probed by future higher signal-to-noise ratio observations.

\subsection{Direct accretion efficiency determination}

\cite{2011ApJ...728...98D} outline how accretion efficiences (hence spins) can be calculated using the monochromatic optical luminosity to determine the mass accretion rate (according to standard accretion disc models) and a suitable determination of the bolometric luminosity from the SED, and determine $\eta$ for 80 Palomar-Green (PG) quasars with black hole mass estimates; their efficiencies are consistent with a large fraction of high-spin objects. \cite{2012MNRAS.419.2529R} employed their method to determine accretion efficiences in various samples of AGN at low and intermediate ($z<1.0$) redshifts, but using a variety of bolometric luminosity estimators.  They find numerous uncertainties are associated with direct determination of $\eta$ such as the choice of accretion disc model, uncertainties in the black hole mass or bolometric luminosity, inclination, dust extinction and host galaxy contamination.  They also find a range of efficiencies, but with a skew towards high spin ($\eta > 0.057$).  \cite{2014ApJ...789L...9T} use the same method to determine the accretion efficiency of 72 luminous unobscured high-redshift ($1.5 < z < 3.5$) AGN, and find compelling evidence for high radiative efficiencies (therefore spins) at high redshift.  They present a full consideration of the various errors that can contribute to efficiency measurements, including some of the uncertainties first outlined in \cite{2012MNRAS.419.2529R}, but still find that the majority of their sample requires high spin ($a>0.7$).

\subsection{Reverberation mapping}

Another independent suggestion of high spin is the broad-line region radius--luminosity relation found from reverberation mapping studies: \cite{2014ApJ...792L..13W} argue that the observed tightness of this $R_{\rm BLR}-L$ relation itself must imply high spin, because the reverberation lag of H$\beta$ to the continuum depends very sensitively on black hole spin; in particular, retrograde accretion in luminous quasars would show strong deviation from the observed $R_{\rm BLR}-L$ relation.  Studying this relationship at different redshifts could in turn lead to a method of estimating black hole spins at different redshifts and studying the evolution of spin, but the observed $R_{\rm BLR}-L$ relation for the current reverberation-mapped AGN itself strongly argues that most of these black holes have rapidly spinning black holes.  This is consistent with an over-representation of rapid spin in the objects that have been used to calibrate the $R_{\rm BLR}-L$ relation.

\subsection{Narrow-Line Seyfert 1 AGN}

Studies of the host galaxies of narrow-line Seyfert 1 AGNs (NLS1s), which constitute a sizeable fraction of nearby AGN (\citealt{2002AJ....124.3042W}, \citealt{2006MNRAS.368..479G}), find that they evolved by secular processes, therefore requiring steady accretion \citep{2011MNRAS.417.2721O} that would again spin up black holes.  The often proposed scenario that NLS1s have high accretion rates and low black hole masses could fit well within this picture since it would imply high radiative efficiency is achieved through high spin.  Finally, high spin is further supported from spectral (e.g., \citealt{2013MNRAS.429.2917F}, \citealt{2015MNRAS.446..633G}) and timing (e.g., \citealt{2010MNRAS.401.2419Z}) observations of indivudal, nearby Narrow-Line Seyfert 1 AGN.

\subsection{Returning radiation and emissivity profiles}

\cite{1976ApJ...208..534C} showed that radiation from the disc can re-radiate the disc itself by light-bending and enhance the reflection fraction, provided the spin is high.  \cite{2002MNRAS.336..315R} found that multiple reflections can increase the amplitude of reflection features. The analysis of \cite{2015MNRAS.446..759C} shows that this may be occurring in IRAS 13224-3809.

\cite{2011MNRAS.414.1269W} showed how the steep emissivity profiles can be generated in situations where light bending is high (i.e., high spin).  The analyses of 1H 0707-495 \citep{2014MNRAS.443.2746W} and Mrk 335 \citep{2015MNRAS.449..129W} provide support for such a scenario.

\subsection{Reflection-dominated quasars at high redshift}

Reflection-dominated X-ray spectral shapes have also been observed at high redshift ($z \approx 2$) and have been identified with Compton-thick or very heavy absorption \citep{2011ApJ...738...44A}.  For those objects which can be fit by pure reflection, the degeneracy between the spectral shapes of Compton-thick AGN and ionised reflection due to light-bending offers scope for a substantial fraction of these sources to have high spin.

\bibliographystyle{mn2e} 
\bibliography{highspinxrb}

\end{document}